\documentstyle[12pt]{article}
\newcommand{\st}{\stackrel}
\newcommand{\eps}{\varepsilon}
\newcommand{\be}{\begin{equation}}

\title{Quadratic brackets from sympletic forms}
\date{July 2, 1993}
\author{Anton Yu.Alekseev
\thanks{On leave from Steklov Mathematical Institute (LOMI),
   Fontanka 27, St.Petersburg 191011, Russia}
\and Ivan T.Todorov
\thanks{On leave from the Institute for Nuclear Research, Bulgarian
  Academy of Sciences, Tsarigradsko Chaussee 72, BG--1784 Sofia, Bulgaria}}
\begin{document}
\begin{center}
\maketitle
Erwin Schr\"{o}dinger International Institute for Mathematical Physics (ESI) \\
Pasteurgasse 6/7, A--1090 Wien, Austria
\end{center}
\vspace{30mm}
\begin{abstract}
We give a physicist oriented survey of Poisson-Lie symmetries of
classical systems.
We consider finite dimensional geometric actions and the chiral WZNW model
as examples for the general construction. An essential point is that quadratic
Poisson brackets appear for group--like variables. It is belived that
after quantization they lead to quadratic exchange algebras.
\vspace{60mm}
\end{abstract}

\pagebreak
\section{Introduction.}
Good old canonical quantization still takes a respectable place among
various quantization schemes. A Hamiltonian approach to the problem
begins with a sympletic 2-form $\Omega$ on a phase space
$\Gamma$. Then one should
invert $\Omega$ and obtain a Poisson bracket (PB)
algebra on $\Gamma$.
Next, one selects a subset of PB to be represented by commutators
upon quantization. We are only aware of a general principle for
such a selection in the case when our physical system admits a
symmetry algebra, that is, a Lie algebra $\cal G$
spanned by a
set of generators
$X_{a}$ satisfying linear PB relations
\begin{equation}
\{X_{a},X_{b}\}=f_{ab}^{c}X_{c}.
\label{1.1}
\end{equation}
The skew symmetry of the structure constants $f$ in the pair of lower
indices and the Jacobi identity
\begin{equation}
f_{ab}^{s}f_{sc}^{d}+f_{bc}^{s}f_{sa}^{d}+f_{ca}^{s}f_{sb}^{d}=0
\label{1.2}
\end{equation}
are necessary and sufficient for the existence of such a Lie algebra.
If a set of variables $Y_{\mu}$ transform linearly under the PB action
of $G$ we require that their covariance properties are preserved
under quantization:
\begin{equation}
\{Y_{\mu},X_{a}\}=f_{\mu a}^{\nu}Y_{\nu}\Rightarrow
[\hat{Y}_{\mu},\hat{X}_{a}]=
i\hbar f_{\mu a}^{\nu}\hat{Y}_{\nu}.
\label{1.3}
\end{equation}
In particular, the PB (\ref{1.1}) goes into a commutation relation for
$\hat{X}_{a}$ with structure constants $i\hbar f_{ab}^{c}$:
\begin{equation}
[\hat{X}_{a},\hat{X}_{b}]=i\hbar \,f_{ab}^{c}\,\hat{X}_c.
\label{1.3a}
\end{equation}
A special covariant family is given by the elements $g$ of
a matrix Lie group
$G$ corresponding to the Lie algebra $\cal G$.
For a basis $\{t_{a} \}$ in the algebra $\cal G$
\begin{equation}
[t_{a},t_{b}]=f_{ab}^{c}t_{c},
\label{1.4}
\end{equation}
left covariance of $g\in G$ under the PB action of $\cal G$
is expressed by
\begin{equation}
\{g,X_{a}\}=t_{a}g.
\label{1.5}
\end{equation} The relation (\ref{1.5}) gives a nonabelian analogue
of the basic PB $\{q,p\}=1$ for coordinate and momentum.
Typically,  group -- like variables $g$ satisfy quadratic PB relations
\cite{1}.
Using the tensor notation
\begin{equation}
g^{1}=g\otimes{\bf 1},\;\;\;g^{2}={\bf 1}\otimes g,
\label{1.6}
\end{equation}
we can write the simplest example of quadratic brackets:
\begin{equation}
\{\stackrel{1}{g},\stackrel{2}{g}\}=\stackrel{1}{g}\stackrel{2}{g}r.
\label{1.7}
\end{equation}
The Jacobi identity for the bracket (\ref{1.7}) is equivalent
to the classical
Yang-Baxter equation for the matrix $r$
\begin{equation}
[r_{12},r_{13}]+[r_{12},r_{23}]+[r_{13},r_{23}]=0.
\label{1.8}
\end{equation}
Here $r_{ij}$ act  in a triple product $V_{1}\otimes V_{2}\otimes V_{3}$
of finite dimensional spaces: $r_{12}=r\otimes \bf 1$, where $\bf 1$ is
the unit operator in $V_{3}\;\;etc$.
Quantization of the PB (\ref{1.7}) leads to  quadratic exchange
relations \cite{2}
\begin{equation}
\stackrel{1}{g}\stackrel{2}{g}=\stackrel{2}{g}\stackrel{1}{g}R\;\;.
\label{1.9}
\end{equation}
Here the matrix $R$ is a solution of the quantum Yang-Baxter equation
\begin{equation}
R_{12}R_{13}R_{23}=R_{23}R_{13}R_{12}
\label{1.10}
\end{equation}
and $R=R(\hbar)\approx 1+i\hbar r$ when $\hbar \rightarrow 0$.
There is a wide class of interesting examples where such $R-$matrices
are known. Nevertheless, it is still a
challenge to find a regular scheme for quantization of quadratic
brackets.

The PB (\ref{1.7}) have an annoying property:
they appear to violate a symmetry under
right shifts, $g \rightarrow gh^{-1}$. Remarkably, this symmetry can be
restored, in some sense, if we admit non trivial PB for parameters $h$
of the shift. This is a generalization of
super-symmetry, where anti-commutative parameters are required.
If we postulate
\begin{equation}
\{\stackrel{1}{h},\stackrel{2}{h}\}=
[r,\stackrel{1}{h}\stackrel{2}{h}],\;\;\;\;
\{\stackrel{1}{g},\stackrel{2}{h}\}=0,
\label{1.11}
\end{equation}
we discover that the formula
(\ref{1.7}) survives after the right shift \cite{3}:
\begin{equation}
\{g^{1}(h^{1})^{-1},g^{2}(h^{2})^{-1}\}=
g^{1}(h^{1})^{-1}g^{2}(h^{2})^{-1}r.
\label{1.12}
\end{equation}
We note that the product of two
consecutive shifts $h=h_{1}h_{2}$ has the same brackets (\ref{1.11})
 provided that $h_{1}$
has zero PB with $h_{2}$:
\begin{equation}
h=h_{1}h_{2},\{\st{1}{h}_{1},\st{2}{h}_{2}\}=0, \;\;
\{\st{1}{h}_{i},\st{2}{h}_{i}\}=
[r,\st{1}{h}_{i}\st{2}{h}_{i}]\Rightarrow\{\st{1}{h},\st{2}{h}\}=
[r,\st{1}{h}\st{2}{h}].
\label{1.13}
\end{equation}
The message of this simple example is that the symmetry group can
carry a nontrivial PB. If the bracket respects the group
multiplication,
the Lie group supplied with such PB is called Poisson-Lie (PL) group.
The PL
symmetry does not preserve the PB if we treat transformation
parameters just as $c$-numbers. We should take into
account the PB of transformation parameters when
we check the symmetry with respect to PL group action.   The notion of PL
symmetry gives us a possibility to find a symmetry in examples where
it seems to be broken.

The Poisson-Lie symmetry in the classical system leads to  the quantum
group symmetry after quantization \cite{3,4}. This observation
motivates our interest to the general theory of PL symmetry and
to  concrete physical systems enjoing this type of symmetry.

Suppose we have a physical system defined by the action principle
\begin{equation}
S=\int {\cal L}(x,\dot{x})dt.
\label{1.14}
\end{equation}
If we suspect that the transformation $x\rightarrow x+\delta_{\eps}x$
is a symmetry of the system, we can easily check it. The transformation
$\delta_{\eps}$ should preserve the action $S$ up to boundary terms.
In the Section 2 we give a similar criterium \cite{5} how to
recognize  a PL symmetry starting from the action principle.

Usually the PL symmetry appears in systems with quadratic PB. To
get experience of calculations with quadratic brackets we start in the
Section 3 by consideration of the model

\begin{equation}
S=\int tr(i\,P\,u^{-1}\,\dot{u}-\frac{1}{2}P^{2})dt.
\label{1.15}
\end{equation}
Here $u$ is an element of compact semi-simple Lie group $G$, $P$
belongs  to ${\cal H}_{+}^{*}$, positive Weyl chamber in the dual to
corresponding Cartan subalgebra $\cal H$. In the case of $G=SU(n)$ $P$
is a diagonal matrix with ordered eigenvalues:
$P=diag(p_{1},\ldots ,p_{n}),\;\;p_{1}\geq p_{2}\ldots \geq p_{n}$.
The system enjoys classical symmetry under left shifts while the right
symmetry is broken by the choice of Cartan subalgebra $\cal H$.
The bracket for variables
$u$ appears to be quadratic. This bracket was derived in \cite{6}
for the case
of $SU(2)$  and guessed for an arbitrary group $G$. We fill this gap
and derive the PB for $u$ directly from the action.

The system (\ref{1.15}) possesses a one-parameter family of
deformations (see
formula (\ref{(3.26a)})) so
that there is neither left nor right symmetry
in the deformed system.
However, the left symmetry can be  restored using the notion of PL
symmetry.
The PB for the deformed system were suggested in \cite{6,9}
and the action principle
was proposed in \cite{7},\cite{8}.
We put these two ideas together and derive
quadratic PB from the action principle.
The PL symmetry helps a lot in inverting the symplectic form
(\ref{(3.24)}) which has no transparent symmetries.
Let us mention that from the physical point of view
the action (\ref{1.15}) represents
a classical model for spin, the deformed system corresponds to quantum
group analogue of particle spin.
The actions (\ref{1.15}) and (\ref{(3.26a)}) were discovered in the
framework of geometric quantization. For this reason we call them
geometric action and deformed geometric action respectively.

The infinite dimensional case of a chiral WZNW model, which has to a large
extent inspired the present study, is considered in Section 4 (Relevant steps
in working out the Lagrangian approach to WZNW models are contained
in \cite{8},\cite{10}--\cite{13}.)
The basic variable is the group valued field $g(x)$ on the circle $S^{1}$.
More precisely, $g$ is a multi-valued function on $S^{1}$ satisfying the
quasi-periodicity condition
\begin{equation}
g(x+2\pi)=g(x)M.
\label{1.17}
\end{equation}
The  monodromy matrix $M$ does not depend on $x$ but varies with time.
The symplectic form (\ref{(4.4)})
on the chiral
phase space $\Gamma_{WZ}$ involves an integral term and a discrete piece
depending  on the boundary values $g_{1}=g_{0}$ and $g_{2}=g(2\pi)=g_{1}M$,
reminiscent to the finite dimensional 2-form (\ref{(3.24)}) studied
in Section 3.
After reviewing the basic properties of the form (\ref{(4.4)})
(first introduced in \cite{8})
including an elementary derivation of the condition that $\Omega$ is a closed
form, $d\Omega=0$, we establish a new result: the presence of
an infinite ``quantum symmetry'' of $\Omega$ under right shifts
of $g(x)$ satisfying
certain boundary conditions. However, the $x$-independent
part of this ``Poisson-Lie symmetry'' (corresponding to the standard quantum
group
symmetry noted previously in various
contexts -- cf.\cite{8},\cite{12}--\cite{16}) is sufficient to
explain the cancellation between world sheet and quantum group monodromy,
established earlier \cite{15,16}.

\section{Quantum group symmetry of symplectic form.}
\setcounter{equation}{0}
Our starting point will be a closed 2-form $\Omega$ on a manifold
$\Gamma$.
Suppose that it is non--degenerate.
The condition
\begin{equation}
d\Omega=0
\label{(2.1)}
\end{equation}
implies the existence, at least locally, of a 1-form $\Theta$ such that
\begin{equation}
\Omega=d\Theta,\;\Theta=P_{i}\,dx^{i},\;P_{i}=P_{i}(x).
\label{(2.2)}
\end{equation}
Hence, if we choose a Hamiltonian $H=H(x)$ on $\Gamma$,
we can define a Lagrangian
$\cal L$ and the corresponding action $S$ as:
\begin{equation}
{\cal L} =P_{i}(x)\dot{x}_{i}-H(x),\;S=\int {\cal L} \,dt
=\int (d^{-1}\Omega -H\,dt),
\label{(2.3)}
\end{equation}
where a dot (on $x$) denotes, as usually, a time derivative.

We recall that an infinitesimal transformation $x\to x+\delta_{\eps}x$ of
$\Gamma$ with parameters $\eps^{a}$
\begin{equation}
\delta_{\eps}x=\eps^{a}\delta_{a}x
\label{(2.4)}
\end{equation}
defines a classical symmetry
if

$(i)$ it leaves the Hamiltonian invariant:
\begin{equation}
\delta_{\eps}H=H(x+\delta_{\eps}x)-H(x)=O(\eps^{2}),
\label{(2.5)}
\end{equation}

$(ii)$ it preserves $\Omega$.

It follows from the definition that the action is invariant
(up to boundary terms)
provided that $\eps$ is time independent. If $\eps^{a}=\eps^{a}(t)$
then
\begin{equation}
\delta_{\eps}S=\int X_{a}\dot{\eps}^{a}dt\;\;(+\;\;
boundary\;\;\;terms).
\label{(2.6)}
\end{equation}
Functions $X_{a}$ generate the symmetry via PB:
\begin{equation}
\delta_{a}f=\frac{\partial f}{\partial x}\delta_{a}x=
\{f,X_{a}\}.
\label{(2.7)}
\end{equation}

The set  of functions $X_{a}$ can be unified into the vector
$X=X_{a}t_{a}$. Usually $X$ is treated as an element of the space
$\cal G^{*}$ dual to Lie algebra $\cal G$. It is motivated by the
formula (\ref{(2.6)}) where it appears in the pairing with
the transformation parameter $\eps$. Generators $X_{a}$ form a closed
algebra with respect to the PB:
\begin{equation}
\{X_{a},X_{b}\}=f^{c}_{ab}X_{c}.
\label{(2.7a)}
\end{equation}
In principle the r.h.s. may be modified by the central extension
(also called anomaly) but we ignore this subtlety and restrict
ourselves to the regular situation. The PB (\ref{(2.7a)}) on the space
$\cal G^{*}$ is called Kirillov-Kostant bracket. It provides a
simplest example of the system with built-in group $G$ symmetry. Let
us introduce auxillary variables $P\in \cal H_{+}^{*}$ (see
definition in the Introduction) and a group element $u$ so that
\begin{equation}
X=u\,P\,u^{-1}.
\label{(2.7b)}
\end{equation}
It is remarkable that the elementary PB (\ref{(2.7a)}) follows from
the action
\begin{equation}
S(x)=\int(tr\,i\,P\,u^{-1}\,du-Hdt).
\label{(2.7c)}
\end{equation}
The variable $P$ may be fixed to be a constant or it may take values in
$\cal H_{+}^{*}$. In the first case our system lives on the
conjugancy class in $\cal G^{*}$ (called also coadjoint orbit) and
reproduces PB (\ref{(2.7a)}). For fixed $P$, $u$ is  defined only up
to multiplication by a diagonal factor from the right. In this paper
we choose
the second case, so that $P$ and $u$ are both dynamical variables. The
calculation of PB for $u$ is a subject of the next Section.

Now let us turn to the notion of Poisson-Lie symmetry.
It is convenient to rewrite the formula (\ref{(2.6)}) in the following form
\begin{equation}
\delta_{\eps}S=-\int\eps^{a}A_{a}\;\; (+\;\;boundary\;\;terms),
\label{(2.8)}
\end{equation}
where we have introduced connection 1-forms
\begin{equation}
A_{a}=A_{a}^{i}\,dx^{i}=\frac{\partial X_{a}}{\partial x^{i}}\,dx^{i}.
\label{(2.8a)}
\end{equation}
The connection $A_{a}$ satisfies abelian zero-curvature condition
\begin{equation}
dA_{a}=0.
\label{(2.8b)}
\end{equation}
The Poisson-Lie (PL) symmetry is a generalization of the above
concept but the condition (\ref{(2.8b)}) is replaced by the weaker
condition
\begin{equation}
dA_{a}=F_{a}^{bc}A_{b}A_{c},
\label{(2.9)}
\end{equation}
where the $F_{a}^{bc}$ are $x$-independent structure constants.
They form a skew-symmetric tensor with respect to the
upper indices and must obey the Jacobi identity (\ref{1.2}). Hence,
they can be viewed as
structure constants of the Lie algebra $\cal G^{*}$.

Let $\tau^{a}$ be (say, matrix) generators of $\cal G^{*}$ such that
\begin{equation}
[\tau^{a},\tau^{b}]=F_{c}^{ab}\tau^{c}.
\label{(2.10)}
\end{equation}
Let us define the group element
$g^{*}=g^{*}(x),\;x\in\Gamma$ as a solution
of the equation
\begin{equation}
(d-A_{a}\tau^{a})g^{*}=0.
\label{(2.11)}
\end{equation}
The group -- like variable $g^{*}$ is a substitute for all
generators $X^{a}$. It is easy to check that
\begin{equation}
\delta_{a}f=\frac{\partial f}{\partial x}\,\delta_{a}\,x=
tr(t_{a}\{f,g^{*}\}{g^{*}}^{-1}).
\label{(2.12)}
\end{equation}
The PB  is not invariant
under such generalized ``quantum'' symmetry transformation. Instead,
we have a complicated deviation in the r.h.s.
\begin{equation}
\{f_{1},f_{2}\}(x+\delta_{\eps}x)=
\{f_{1}(x+\delta_{\eps}x),f_{2}(x+\delta_{\eps}x)\}_{x}+
\frac{\partial f_{1}}{\partial x^{i}}
\frac{\partial (\delta_{\eps}x^{i})}{\partial \eps^{a}}
\frac{\partial f_{2}}{\partial x^{j}}
\frac{\partial (\delta_{\eps}x^{j})}{\partial \eps^{b}}
F_{c}^{ab}\eps^{c}.
\label{(2.13)}
\end{equation}
Having ascribed the PB
\begin{equation}
\{\eps^{a},\eps^{b}\}=F_{c}^{ab}\eps^{c}
\label{(2.14)}
\end{equation}
to symmetry parameters $\eps^{a}$
we can restore the symmetry by considering $f(x+\delta_{\eps}x)$
as a function of both $x$ and $\eps$ and including the second term
in the r.h.s. of
(\ref{(2.13)})
into the definition of PB:
\begin{equation}
\{f_{1},f_{2}\}(x+\delta_{\eps})=
\{f_{1}(x+\delta_{\eps}x),f_{2}(x+\delta_{\eps}x)\}_{x}+
\{f_{1}(x+\delta_{\eps}x),f_{2}(x+\delta_{\eps}x)\}_{\eps}.
\label{(2.14a)}
\end{equation}

Now  we shall discuss a basic example of the algebra $\cal G^{*}$
useful for the next Sections.

Let $t_{a}$ be the matrix generators of a Lie algebra $\cal G$ satisfying
(\ref{1.4}).
Assuming that $\cal G$ is semi-simple we choose a Cartan subalgebra with
basis $t_{j}=H_{j}\;(1\leq j\leq rank\,\cal G)$ and split the remaining
generators into $t_{\alpha}$ and $t_{-\alpha}$ where $\alpha$ enumerates
positive roots. Setting
\begin{equation}
\eps=\eps^{a}t_{a}
\label{(2.15)}
\end{equation}
we define a PB between commuting matrices
\begin{equation}
\st{1}{\eps}=\eps\otimes 1 \;\; and \;\; \st{2}{\eps}=1\otimes\eps
\label{(2.16)}
\end{equation}
by the classical $r$-matrix relation \cite{1}
\begin{equation}
\{\st{1}{\eps},\st{2}{\eps}\}=[r,\st{1}{\eps}+\st{2}{\eps}],
\label{(2.17)}
\end{equation}
where $r$ is skew-symmetric:
\begin{equation}
r=-\sum_{\alpha >0}
(t_{\alpha}\otimes t_{-\alpha}-t_{-\alpha} \otimes t_{\alpha}).
\label{(2.18)}
\end{equation}
We thus arrive at a new Lie algebra $\cal G^{*}$ with structure constants
$F_{ab}^{c}$ related to those of $\cal G$ by the following formulae:

\begin{eqnarray}
F_{c}^{\alpha\beta} & = &
f_{\alpha\beta}^{c}\;\;\;\;\;\;\;\;\;\;\;\;\;
\mbox{\it{if}} \;\;\; \alpha > 0,\beta>0, \nonumber \\
F_{c}^{\alpha\beta} & = & -f_{\alpha\beta}^{c}\;\;\;\;\;\;\;\;\;\;\;
\mbox{\it{if}} \;\;\; \alpha < 0,\beta<0, \nonumber \\
F_{\alpha}^{\alpha i} & = &
f_{\alpha i}^{\alpha}\;\;\;\;\;\;\;\;\;\;\;\;\;\;
\mbox{\it{if}} \;\;\; \alpha > 0,
\\
F_{\as
\beginlpha j} & = &
f_{\alpha j}^{\alpha}\;\;\;\;\;\;\;\;\;\;\;\;\;\;
\mbox{\it{if}} \;\;\; \alpha < 0, \nonumber \\
F_{c}^{\alpha\beta} & = & -F^{\alpha\beta}_{c}=0\;\;\;
\mbox{\it{if}} \;\;\; \alpha > 0,\beta<0. \nonumber
\label{(2.19)}
\end{eqnarray}
Here, as before, $i,j$ label  Cartan generators of $\cal G$.
Lie algebra $\cal G^{*}$
can be embedded into the direct sum of two Borel subalgebras of $\cal G$
spanned by $(t_{\alpha},H_{i})$ and by $(t_{-\alpha},H_{i})\;(\alpha > 0)$
respectively. More precisely, a pair $(b_{+},b_{-})$ belongs to $\cal G^{*}$
if the matrices $b_{\pm}$ have opposite diagonal parts:
\begin{equation}
(b_{+},b_{-})\in {\cal G^{*}} \;\;\;
\mbox{\it{if}} \;\;\; b_{+}^{diag}+b_{-}^{diag}=0.
\label{(2.20)}
\end{equation}
The pair $(b_{+},b_{-})$ can be parametrized by the element
$b=b_{+}-b_{-}$ in the algebra $\cal G$.

We shall also use a group $G^{*}$ with Lie algebra $\cal G^{*}$. It is
contained in the product of Borel subgroups of $G$ (the subgroups of
upper and lower triangular matrices in the case of SL $(n,C)$):
\begin{equation}
(L_{+},L_{-})\in G^{*}
\;\;\;
\mbox{\it{if}} \;\;\; L_{+}^{diag}L_{-}^{diag}=1.
\label{(2.21)}
\end{equation}

The product in $G^{*}$ is defined component wise:
\begin{equation}
(L_{+},L_{-})(M_{+},M_{-})=(L_{+}M_{+},L_{-}M_{-}),
\label{(2.22)}
\end{equation}
while the connection form $A=dg^{*}g^{*-1}$ is given by
\begin{equation}
A=(dL_{+}L_{+}^{-1},dL_{-}L_{-}^{-1}).
\label{(2.23)}
\end{equation}
Embedded into the algebra $\cal G$ the connection acquires  the form
\begin{equation}
A=dL_{+}L_{+}^{-1}-dL_{-}L_{-}^{-1}.
\label{(2.23a)}
\end{equation}
It is important to note that the PB (\ref{(2.17)})
for infinitesimal matrices can be
exponentiated resulting in Sklyanin's quadratic algebra \cite{1} for
(matrix) group elements (cf.(\ref{1.11})),
\begin{equation}
\{\st{1}{v},\st{2}{v}\}=[r,\st{1}{v}\st{2}{v}],
\label{(2.24)}
\end{equation}
where we have used again  tensor notation (\ref{1.6}). The global
version of the formula  (\ref{(2.14a)}) looks like follows:
\begin{equation}
\{f_{1},f_{2}\}(x^{v})=
\{f_{1}(x^{v}),f_{2}(x^{v})\}_{x}+
\{f_{1}(x^{v}),f_{2}(x^{v})\}_{v},
\label{(2.25)}
\end{equation}
where $x^{v}$ is a point $x$ shifted by the group element $v$.

\section{Geometric action and its quantum group
deformation.}
\setcounter{equation}{0}

Let $G$ be a simple compact Lie group with an ordered Cartan-Weyl
basis of its Lie algebra. We define the (left) chiral phase space
$\Gamma=\Gamma(G,\cal H)$ as a Cartesian product of $G$ and
$\cal H^{*}_{+}$ (see definition in Section 1).
The (undeformed) classical Lagrangian, invariant under left shifts of
$u\in G$, can be read from the action (\ref{1.15}):
\begin{equation}
{\cal L} = tr(iPu^{-1}\dot{u}-\frac{1}{2}P^{2}).
\label{(3.1)}
\end{equation}
The resulting system has been proposed \cite{17} as a finite
dimensional analogue of a chiral WZNW model. The corresponding
quantum mechanical state space contains every finite dimensional
irreducible representation of $G$ with multiplicity equal to one.

Infinitesimal left shifts
\begin{equation}
\delta_{\eps}u=\eps u,\;\;\delta_{\eps}P=0
\label{(3.2)}
\end{equation}
satisfy conditions $(i)$ and $(ii)$ of Section 2. If ${t_{a}}$ is a
basis in $\cal G$ so that
\begin{equation}
\eps =\eps^{a}t_{a},
\label{(3.3)}
\end{equation}
then symmetry generators $X_{a}$
\begin{equation}
X_{a}=tr(t_{a}\,u\,P\,u^{-1})
\label{(3.4)}
\end{equation}
appear in the variation of the action (\ref{(2.6)}).
The symplectic form $\Omega_{0}$ associated to the Lagrangian
(\ref{(3.1)}) can be written either in terms of
$P$ and $u^{-1}du$
\begin{equation}
\Omega_{0}=i\,tr\,(dP\,u^{-1}\,du-P(u^{-1}\,du)^{2}),
\label{(3.5a)}
\end{equation}
or as a bilinear form on $\cal G^{*}\times\cal G$,
\begin{equation}
\Omega_{0}=\frac{i}{2}tr(dl\,du\,u^{-1}+dP\,u^{-1}\,du),
\label{(3.5b)}
\end{equation}
where
\begin{equation}
l=u\,P\,u^{-1}.
\label{(3.6)}
\end{equation}
The equivalence between (\ref{(3.5a)}) and (\ref{(3.5b)}) is  established
using the identity
\begin{equation}
dl=[du\,u^{-1},l]+u\,dP\,u^{-1}.\;\;
\label{(3.6a)}
\end{equation}
Here and in what follows we omit the wedge sign $\wedge$ in the
products of exterior differentials.

In order to bring $\Omega_{0}$ to a canonical form we first introduce
a basis of left invariant 1-forms $\Theta^{a}$ on $G$ setting
\begin{equation}
i\,u^{-1}du=\Theta^{a}t_{a}=
\sum_{j}\Theta^{j}H_{j}+\sum_{\alpha >0} (\Theta^{-\alpha}t_{-\alpha}+
\Theta^{\alpha}\,t_{\alpha}),
\label{(3.7)}
\end{equation}
where the sum in $\alpha$ runs over the positive roots of $\cal G$,
while the first term involves a sum (in $j$) over the basis in Cartan
subalgebra.
In the special case ${\cal G} =SL(n)$ we have
\begin{equation}
t_{\alpha}= e_{jk}\;j<k,\;\;
t_{-\alpha}=e_{kj}=e_{jk}^{*},\;\;
H_{j}=e_{jj}-e_{j+1\,j+1},\;1\leq j\leq n-1,
\label{(3.8a)}
\end{equation}
where $e_{jk}$ are  $n$ by $n$ Weyl matrices characterized by
the product formula
\begin{equation}
e_{ij}e_{kl}=\delta_{jk}e_{il}.
\label{(3.8b)}
\end{equation}
More generally, the basis in $\cal G$ is characterized by the
commutation relations
\begin{equation}
[H_{j},t_{\pm \alpha}]=
\pm 2\frac{<\alpha_{j},\alpha>}{\alpha_{j}^{2}}t_{\pm \alpha},\;
[t_{\alpha_{j}},t_{-\alpha_{j}}]=H_{j},
\label{(3.9)}
\end{equation}
where $\alpha_{j}\; (j=1,\cdots,rank\;\cal G)$ are the simple roots
of $\cal G$. We can expand $P$ using
the basis of fundamental weights $\{h^{i}\}$ dual to $\{H_{i}\}$:
\begin{equation}
P=P_{j}h^{j}(\in {\cal H_{+}^{*}}),
\label{(3.10)}
\end{equation}
where
\begin{equation}
(h^{i},H_{j})=tr\,h^{i}\,H_{j}=\delta_{j}^{i}.
\label{(3.11)}
\end{equation}
In the case  of $SL(n)$ we have
\begin{equation}
h^{j}=e_{jj}-\frac{1}{n}{\bf 1}\;({\bf 1}=\sum_{i=1}^{n}e_{ii}).
\label{(3.12)}
\end{equation}
Setting further
\begin{equation}
[t_{\alpha},t_{-\alpha}]=H_{\alpha},
\label{(3.13a)}
\end{equation}
\begin{equation}
(P,H_{\alpha})=P_{\alpha}
\label{(3.13b)}
\end{equation}
we transform (\ref{(3.5a)}) into
\begin{equation}
\Omega_{0}= \sum_{j}
dP_{j}\Theta^{j}+i\sum_{\alpha}P_{\alpha}\Theta^{\alpha}\Theta^{-\alpha}.
\label{(3.14)}
\end{equation}

The canonical expression (\ref{(3.14)}) for $\Omega_{0}$ has the advantage of
being readily invertible. To write the corresponding PB we introduce
vector fields $V_{i}$ and $V_{\pm \alpha}$ dual to $\Theta^{i}$
and $\Theta_{\pm\alpha}$ (in the sense that
$<\Theta^{a},V_{b}>=\delta_{b}^{a}$):
\begin{eqnarray}
V_{i}=-i\,tr\,u\,H_{i}\frac{\partial}{\partial u},\nonumber \\
V_{\pm \alpha}=-i\, tr\,u\,t_{\pm \alpha} \frac{\partial}{\partial u}.
\label{(3.16)}
\end{eqnarray}
The resulting PB is given by the skew product
\begin{equation}
{\cal P} =\sum_{j}V_{j}\wedge\frac{\partial}
{\partial P_{j}}+i\sum_{\alpha >0}
\frac{1}{P_{\alpha}}V_{\alpha}\wedge V_{-\alpha}.
\label{(3.17a)}
\end{equation}
If $f_{1,2}$ are two functions of $P$ and $u$, their PB can be
calculated as
\begin{equation}
\{f_{1},f_{2}\}={\cal P} (f_{1},f_{2}).
\label{(3.17b)}
\end{equation}
In particular, the non-vanishing PB for the basic phase space
variables $P$ and $u$ are
\begin{equation}
\{\st{1}{u},\st{2}{u}\}=-\st{1}{u}\st{2}{u}\,r_{0}(P),
\label{(3.18)}
\end{equation}
\begin{equation}
\{\st{1}{u},\st{2}{P}\}=-i\,\st{1}{u}\rho.
\label{(3.19)}
\end{equation}
Here $r_{0}$ is an $r-$matrix depending on the dynamical variables
$P_{\alpha}$:
\begin{equation}
r_{0}(P)=
\sum_{\alpha}
\frac{i}{P_{\alpha}}
(t_{\alpha}\otimes t_{-\alpha}-t_{-\alpha}\otimes t_{\alpha}),
\label{(3.20)}
\end{equation}
while
\begin{equation}
\rho=\sum_{j} H_{j} \otimes h^{j}.
\label{(3.21)}
\end{equation}
The matrix $r_{0}(p)$ can be regarded as a classical analogue of
$6j$-symbol (see \cite{6} for details).

{\bf Remark.} The linear relation (\ref{(3.19)}) is readily
quantized:
\begin{equation}
[\st{1}{u},\st{2}{P}]=\hbar \st{1}{u}\rho.
\label{(3.22)}
\end{equation}
The quantum $R-$matrix corresponding to (\ref{(3.18)}) was proposed in
\cite{6}. It would be interesting to work out a general algorithm
permitting to derive it starting from the classical action principle.

Now we turn to our main example, the quantum group deformation of
the model (\ref{(3.1)}).
Let us introduce the exponentiated angular momentum matrix
$L$ and its Gauss decomposition:
\begin{equation}
L=ue^{i\gamma P}u^{-1}=L_{+}L_{-}^{-1}.
\label{(3.23)}
\end{equation}
Here $L_{\pm}$ belong to the Borel subgroups $B_{\pm}$ of $G$
generated by $H_{j}$ and $t_{\pm\alpha}$ and product of their
diagonal parts is equal to one -- cf.(\ref{(2.21)}).

The deformed counterpart of the symplectic form (\ref{(3.5a)}) looks
as follows \cite{8,7}:

\begin{equation}
\Omega_{\gamma}=tr\{
dP\,i\,u^{-1}\,du-
\frac{1}{2\gamma}
(e^{i\gamma P}u^{-1}\,du\,
e^{-i\gamma P}\,u^{-1}\,du
+L_{+}^{-1}dL_{+}L_{-}^{-1}dL_{-})
\}.
\label{(3.24)}
\end{equation}
In verifying that $\Omega_{\gamma}$ is closed we use
the relations
\begin{eqnarray}
\frac{1}{6\gamma}tr(L^{-1}dL)^{3}=
-\frac{1}{6\gamma} tr(L_{-}^{-1} dL_{-} - L_{+}^{-1} dL_{+})^{3}=
\frac{1}{2\gamma} d(tr L_{+}^{-1} dL_{+} L_{-}^{-1} dL_{-});
\label{(3.24a)}
\end{eqnarray}
\begin{eqnarray}
\frac{1}{6\gamma}tr(L^{-1}dL)^{3}
=tr\{
dP\,i\,u^{-1}\,du-
\frac{1}{2\gamma}
e^{i\gamma P}u^{-1}\,du\,
e^{-i\gamma P}\,u^{-1}\,du\}.
\label{(3.25)}
\end{eqnarray}

Let us mention that the  formula (\ref{(3.24a)}) is exactly the
same as we use to evaluate  $d^{-1}\,tr(dg\,g^{-1})^{3}$ in the WZNW
model. The representation of the chiral field $g$ as a
product of Gauss components leads to bosonisation of WZNW
action.

It is clear from the equation
\be
tr\{
(idP-\frac{1}{2\gamma}
e^{i\gamma P}\,u^{-1}\,du\,e^{i\gamma P})
u^{-1}du\}=
dP_{j}\Theta^{j}+i
\sum_{\alpha >0}
\frac{\sin \gamma P_{\alpha}}{\gamma}
\Theta^{\alpha}\Theta^{-\alpha}.
\label{(3.26)}
\end{equation}
that the first two terms in
(\ref{(3.24)})
reproduce the undeformed
expression (\ref{(3.14)}) in the limit $\gamma \rightarrow 0$.
As we shall see shortly, the third term vanishes
when $\gamma \rightarrow 0$ (in accord with (\ref{(3.25)})).
The deformed geometric action can be written as
\begin{equation}
S_{\gamma}=\int (d^{-1}\,
\Omega_{\gamma}(u,p)-\frac{1}{2}\,tr\,P^{2}\,dt).
\label{(3.26a)}
\end{equation}
To write an explicit form for the action one must solve the
equation
\begin{equation}
d\alpha_{\gamma}(u,P)=\Omega_{\gamma}.
\label{(3.26b)}
\end{equation}
We were not able to find a nice expression for $\alpha_{\gamma}$.
Fortunately, boundary terms don't contribute in
PB and we can leave the
deformed geometric action in the form (\ref{(3.26a)}).

The symplectic form (\ref{(3.24)}) possesses a ``quantum'' symmetry under left
shifts
$\delta_{\eps}u=\eps u$, $\delta_{\eps}P=0$:
\be
\delta_{\eps}\Omega_{\gamma}=
-d\,tr(\eps(t)(dL_{+}L_{+}^{-1}-dL_{-}L_{-}^{-1})).
\label{(3.27)}
\end{equation}
The derivation of the formula (\ref{(3.27)}) is straightforward but
quite long. We suggest it as an exercise for interested reader (see also
\cite{7}, where formula (\ref{(3.27)}) is proved in more general
setting). At this point  we unravel the PL symmetry in the system.
Naively, the deformed form $\Omega_{\gamma}$ is neither left nor right
invariant. The formula (\ref{(3.27)}) shows that the system enjoys a
PL symmetry with respect to left shifts.

Now we proceed to computing the PB of the dynamical variables on
$\Gamma$. The PB of $L_{\pm}$ are easier to find -- see \cite{8,7}. The
calculation of the PB for a pair of $u$'s requires some work
(although their expression has been conjectured correctly in
\cite{6}). We shall first compute $\{\st{1}{u},\st{2}{u}\}$ at the
group unit and then will use the technique  of Section 2 to
derive the PB for arbitrary $u$'s.

For $u=1$ the right invariant form $i\,du\,u^{-1}$
coincides with the left invariant one (\ref{(3.7)}) while $L_{\pm}
= e^{\pm i \frac{\gamma}{2} P}$. We find
\begin{eqnarray}
iL_{+}^{-1}dL_{+}=-\frac{\gamma}{2}dP
-2i \sum_{\alpha >0} t_{\alpha}\Theta^{\alpha}
\sin \frac{\gamma}{2}P_{\alpha}, \nonumber \\
iL_{-}^{-1}dL_{-}=\frac{\gamma}{2}dP-2i
\sum_{\alpha >0}t_{-\alpha}\Theta^{-\alpha}
\sin \frac{\gamma}{2}P_{\alpha}
\label{(3.28)}
\end{eqnarray}
and hence we can calculate the symplectic structure at the group unit
\be
\Omega_{\gamma}\mid_{u=1}=\sum_{j}dP_{j}\Theta^{j}+
\frac{1}{\gamma}
\sum_{\alpha >0}
(e^{i\gamma P_{\alpha}}-1)
\Theta^{\alpha}\Theta^{-\alpha}.
\label{(3.29)}
\end{equation}
This gives a counterpart of the PB form (\ref{(3.17a)})
\be
{\cal P}_{\gamma}\mid_{u=1}
=\sum_{j}V_{j}\wedge\frac{\partial}{\partial P_{j}}
+\sum_{\alpha}\frac{\gamma}
{1-e^{i\gamma P_{\alpha}}}V_{\alpha}\wedge V_{-\alpha}.
\label{(3.30)}
\end{equation}
Now the PB of $u$ and $P$ at the group unit can be written:
\be
\{\st{1}{u},\st{2}{P}\}\mid_{u=1}=-i\,\rho,
\;\;\;
\{\st{1}{u},\st{2}{u}\}\mid_{u=1}=\tilde{r}_{\gamma}(P),
\label{(3.31)}
\end{equation}
where
\be
\tilde{r}_{\gamma}(P)=
\sum_{\alpha >0}
\frac{\gamma}{e^{i\gamma P_{\alpha}}-1}
(t_{\alpha}\otimes t_{-\alpha}-t_{-\alpha}\otimes t_{\alpha}).
\label{(3.32)}
\end{equation}

To evaluate the PB for arbitrary $u$ we use the ``quantum'' symmetry
under left shifts and Eq.(\ref{(2.25)}):
\begin{eqnarray}
\{\st{1}{u},\st{2}{u}\}
=\{v^{1}u^{1},v^{2}u^{2}\}^{u}
\mid_{v=u,u=1}+
\{v^{1}u^{1},v^{2}u^{2}\}^{v}
\mid_{v=u,u=1}=  \nonumber \\
=\st{1}{v}\st{2}{v}\,\tilde{r}_{\gamma}(P)\mid_{v=u}+
\gamma[r,\st{1}{v}\st{2}{v}]\mid_{v=u}= \nonumber \\
=\st{1}{u}\st{2}{u}\,\tilde{r}_{\gamma}(P)+\gamma[r,\st{1}{u}\st{2}{u}]=
\gamma r\st{1}{u}\st{2}{u}+\st{1}{u}\st{2}{u}
r_{\gamma}(P).
\label{(3.33)}
\end{eqnarray}
where $r$ is the classical skew symmetric $r-$matrix (\ref{(2.18)})
and $r_{\gamma}(P)$ is equal to
\be
r_{\gamma}(P)=\tilde{r}_{\gamma}-\gamma r=
\sum_{\alpha >0}
\frac{\gamma e^{i\gamma P_{\alpha}}}{e^{i\gamma P_{\alpha}}-1}
(t_{\alpha}\otimes t_{-\alpha}-t_{-\alpha}\otimes t_{\alpha}).
\label{(3.34)}
\end{equation}
The PB of $u$ and $P$ just reproduces
(\ref{(3.19)}). Restored left symmetry enabled us to reduce the
calculation of complicated quadratic PB (\ref{(3.33)}) to the group
unit. Let us remark that we computed the PB algebra for the classical
geometric action by means of the same trick. Using  left-invariant
Maurer-Cartan forms we reduced the problem to the group unit and
calculated the Poisson bivector (\ref{(3.17a)}). Thus, PL symmetry
appears to be as powerful as a usual one but it could be applied
to systems where the symmetry is broken in the special way.

\section{Chiral WZNW model.}
\setcounter{equation}{0}

The phase space $\Gamma_{WZ}$ of chiral WZNW model is spanned by
quasi-periodic $G-$valued functions $g$ on the circle ${\bf S^{1}}$
satisfying quasi-periodic boundary conditions (\ref{1.17}) with
$x-$independent monodromy matrix $M$. One expresses the
symplectic form on $\Gamma_{WZ}$ in terms of the left invariant form
$g^{-1}(x)\,dg(x)$, the boundary values
\begin{equation}
g_{1}=g(0),\;\;\; g_{2}=g(2\pi )
\label{(4.1)}
\end{equation}
and the Gauss decomposition of the monodromy
\begin{equation}
M=g_{1}^{-1}g_{2}=M_{+}M_{-}^{-1},
\label{(4.2)}
\end{equation}
where we require that the diagonal parts of $M_{\pm}$ cancel each
other (as in the case of $L_{\pm}$),
\begin{equation}
M_{+}^{diag}M_{-}^{diag}=1.
\label{(4.3)}
\end{equation}
The symplectic form on $\Gamma_{WZ}$ is written as \cite{8}
\begin{equation}
\Omega=\frac{1}{2\gamma}tr
\{
\int_{0}^{2\pi}g^{-1}\,dg\partial (g^{-1}\,dg)dx+
dg_{1}\,g_{1}^{-1}\,dg_{2}\,g_{2}^{-1}-
M_{+}^{-1}\,dM_{+}\,M_{-}^{-1}\,dM_{-}
\},
\label{(4.4)}
\end{equation}
where $\gamma=-\frac{4\pi}{k}$ ($k$  is the level in the classical
case and the height, -- i.e., the level plus the dual Coxeter number
-- in the quantum case). The relative coefficients of three
terms in (\ref{(4.4)}) are in fact determined by the requirement that
the $2-$  form is closed,
\begin{equation}
d\Omega =0.
\label{(4.5)}
\end{equation}
In proving (\ref{(4.5)}) we use the following easily verifiable
identities
\begin{equation}
tr\,
\int_{0}^{2\pi}
g^{-1}\,dg\,\partial (g^{-1}\,dg)dx=
\frac{1}{3}
tr\{
(g_{2}^{-1}\,dg_{2})^3-
(g_{1}^{-1}\,dg_{1})^{3}
\};
\label{(4.6a)}
\end{equation}

\begin{equation}
tr\,d(M_{+}^{-1}\,
dM_{+}\,
M_{-}^{-1}\,
dM_{-})=
\frac{1}{3}tr(M^{-1}\,dM)^{3}.
\label{(4.6b)}
\end{equation}

The form (\ref{(4.4)}) corresponds to the left chiral sector of the
WZNW model coupled to the deformed geometric model considered in the
Section 3.
It is straightforward to check that $\Omega$ is invariant under
periodic left shifts
\begin{equation}
\delta_{\eps}g=\eps (x)g(x),\;\;\; \eps(0)=\eps(2\pi).
\label{(4.7)}
\end{equation}

We shall demonstrate that the form (\ref{(4.4)}) also possesses an
infinite ``quantum symmetry'' under right shifts:
\begin{equation}
\delta_{\eps}g(x)=g(x)\eps(x),
\label{(4.8)}
\end{equation}
where $\eps$ is either $x$ independent (the known case of a finite
quantum group symmetry) or satisfies the boundary conditions
\begin{equation}
\eps (0)=\eps_{+}\equiv\eps^{i}H_{i}+
\sum_{\alpha >0}
\eps^{\alpha}t_{\alpha},\;\;\;
\eps (2\pi)=\eps_{-}\equiv
-\eps^{i}H_{i}+
\sum_{\alpha >0}\eps^{-\alpha}t_{\alpha}.
\label{(4.9)}
\end{equation}
In verifying the invariance of the form (\ref{(4.4)}) we use the
following relations:
\begin{equation}
\delta_{\eps}(g^{-1}\,dg)=[g^{-1}\,dg,\eps]+d\eps,
\label{(4.10)}
\end{equation}

\begin{eqnarray}
\delta_{\eps}tr\int g^{-1}\,dg\,\partial (g^{-1}\,dg)dx=
tr\{
2d\int \eps\,\partial (g^{-1}\,dg)dx+       \nonumber \\
+g_{2}^{-1}\,dg_{2}
(d\eps_{-}+[g_{2}^{-1}\,dg_{2},\eps_{-}])
-g_{1}^{-1}\,dg_{1}(d\eps_{+}+[g_{1}^{-1}\,dg_{1},\eps_{+}])
\};
\label{(4.11)}
\end{eqnarray}

\begin{eqnarray}
\delta_{\eps}tr
(dg_{1}\,g_{1}^{-1}\,dg_{2}\,g_{2}^{-1})= \\
=tr(d\eps_{+}g_{1}^{-1}\,dg_{2}\,g_{2}^{-1}\,g_{1}-
d\eps-d\eps_{-} g_{2}^{-1}\,dg_{1}\,g_{1}^{-1}g_{2}); \nonumber
\label{(4.12)}
\end{eqnarray}

\begin{equation}
\delta M_{\pm}=-\eps_{\pm}M_{\pm},\;\;\;
tr(dM_{-}\,M_{-}^{-1}\,d\eps_{-}-
dM_{+}\,M_{+}^{-1}\,d\eps_{+})=0
\label{(4.13a)}
\end{equation}
\begin{equation}
\delta_{\eps}\,tr
(M_{+}^{-1}\,dM_{+}\,M_{-}^{-1}\,dM_{-})=
-tr(M^{-1}\,dM\,d\eps_{-}+
dM\,M^{-1}\,d\eps_{+}).
\label{(4.13b)}
\end{equation}
The result for the sum of three terms is short and elegant
\begin{equation}
\delta_{\eps}\Omega =
2tr\,d\{
\int \eps\,\partial(g^{-1}\,dg)dx+
\eps_{+}\,g_{1}^{-1}\,dg_{1}-
\eps_{-}\,g_{2}^{-1}\,dg_{2}
\}.
\label{(4.14)}
\end{equation}
The r.h.s. of (\ref{(4.14)}) is expressed in terms of Maurer-Cartan
forms $g^{-1}\,dg$.
The formula (\ref{(4.14)}) shows that the chiral WZNW symplectic form
$\Omega$ satisfies conditions of the Section 2 and thus the model
enjoys PL symmetry. It is known \cite{8},\cite{11}-\cite{12} that the
brackets for the field $g$ are quadratic
as in the finite-dimensional example of Section 3.

\vspace{10mm}
{\bf Acknowledgments}

\vspace{7mm}

The   authors thank the Erwin Schr\"{o}dinger International Institute
for Mathematical Physics for hospitality and support during the
course of this work. I.T. acknowledges partial support by the
Bulgarian Science Foundations for Scientific Research under contract
F11.

\end{document}